\documentclass{aa}  

\usepackage{xcolor}
\usepackage{graphicx}
\usepackage{txfonts}

\newcommand{\msun}{M$_{\sun}$}

\newcommand{\vflat}{V$_{\rm flat}$}
\newcommand{\vmax}{V$_{\rm max}$}
\newcommand{\wfif}{W$_{\rm 20}$}
\newcommand\tiri{{\sc TiRiFiC}}
\newcommand\tirific{{\sc TiRiFiC}}
\newcommand\HI{\ion{H}{i}}
\newcommand\hi{\ion{H}{i}}

\newcommand\kms{km s$^{-1}$}

\newcommand{\mhi}{${\rm M_{\ion{H}{i}}}$}

\begin{document}

   \title{The MeerKAT Fornax Survey V.} \subtitle{\HI\ kinematics and Fornax cluster membership of the dwarf galaxy ESO 358-60.}

   \author{P. Kamphuis \inst{1}
          \and
          P. Serra \inst{2}
          \and 
          D. Kleiner \inst{3}
          \and
          R.-J. Dettmar \inst{1}
          \and 
          G. I. G. J\'ozsa \inst{4,5}
            }
\offprints{P. Kamphuis, \email{peter.kamphuis@astro.rub.de}} 
\institute{Ruhr University Bochum, Faculty of Physics and Astronomy, Astronomical Institute (AIRUB), 44780 Bochum, Germany
\and
INAF - Osservatorio Astronomico di Cagliari, Via della Scienza 5, I-09047 Selargius (CA), Italy 
\and
Netherlands Institute for Radio Astronomy (ASTRON), Oude Hoogeveensedijk 4, 7991 PD Dwingeloo, The Netherlands
\and
Max-Planck-Institut f\"ur Radioastronomie, Auf dem Hügel 69, 53121 Bonn
\and
Department of Physics and Electronics, Rhodes University, Artillery Road , Makhanda 1640, South Africa\\
\\
}           

   \date{}

 
  \abstract
   {The MeerKAT Fornax Survey (MFS) is a large survey project mapping the \hi\ in the Fornax cluster. Most of the cluster members detected in \hi\ show significant signs of interaction with the intra-cluster medium or other galaxies. The galaxy ESO\,358-60 however stands out as its  large \hi\ disk appears regular and undisturbed. Combined with the fact that the galaxy's systemic velocity is at the edge of the velocity distribution of Fornax, a possible explanation for this undisturbed disk is that the galaxy is not in Fornax. }
 {Our goal is to understand the detailed morphology and kinematics of the \hi\ disk of ESO\,358-60 and, by doing so, establish whether the galaxy is a member of Fornax.} 
 {We analyze the \hi\ distribution within and around ESO 358-60 based on the MFS observations in a 2 deg$^2$ field around the galaxy. We visually inspect the low resolution data in order to study the \hi\ disk from the center to its outskirts and look for low column density gas that could reveal recent interactions. We then construct a detailed parameterization of the \hi\ disk by fitting a tilted ring model to the high resolution data cube. Using a bootstrap method we establish accurate errors on our best fit models. We use the fitted rotational velocity to place the galaxy on the baryonic Tully-Fisher relationship. By equating the galaxy's \hi\ and 3.6 $\mu$m fluxes to the thus retrieved baryonic mass, we obtain a redshift independent distance.}
   {We confirm that the immediate surroundings of ESO\,358-60 are quiescent relative to other MFS detections and find no obvious interacting companion to the galaxy. Our modeling confirms the regularity of the \hi\ disk in ESO\,358-60 but also shows that the galaxy contains a significant line of sight warp and contains radial motions, of the order of 10 \kms, that cover the extent of the optical disk. From the modeling we obtain a velocity \vflat\ = 48.1 $\pm$ 1.4 \kms\ for the best fit rotation curve. This leads to a distance from the baryonic Tully-Fisher relation of 9.4 $\pm$ 2.5 Mpc, $\sim$ 10 Mpc less than the distance to the Fornax cluster. This distance not only fits better with \vflat\ but also with the overall rotation curves and \hi\ content of low mass galaxies and the fact that the galaxy appears undisturbed and reasonably symmetric. It is also consistent with the distance calculated in the Cosmicflows project.  At 9.4 Mpc ESO\,358-60 cannot be a member of the Fornax cluster but is a foreground field galaxy.}
   {}

   \keywords{ISM: structure --
        Galaxies: evolution --
        Galaxies: intergalactic medium --
        Galaxies: star formation
               }

   \maketitle
%

\section{Introduction}
\begin{figure*}[htbp!]
\hspace*{-1.5cm}\includegraphics[width=2.25\columnwidth]{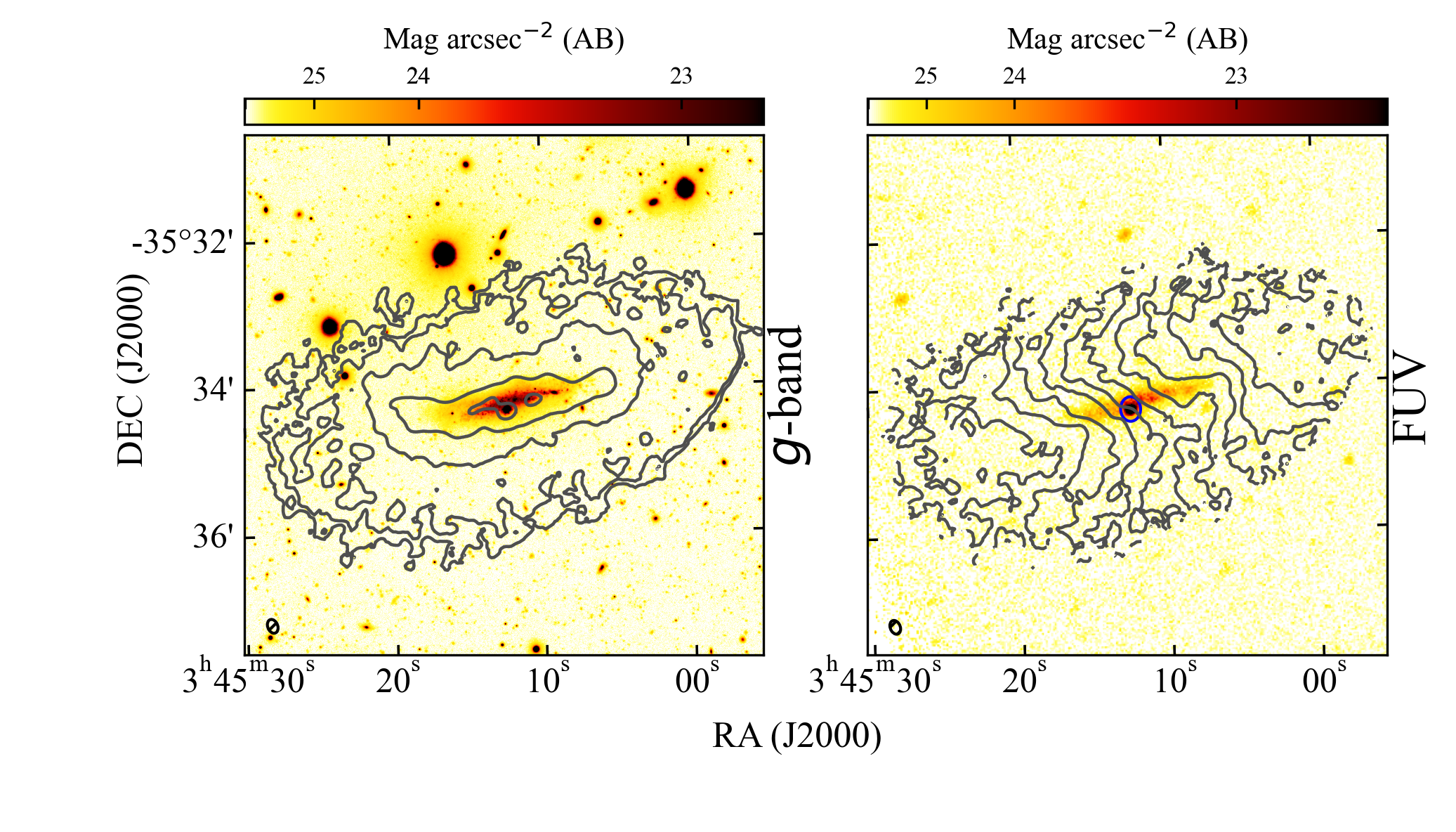}
\caption{Left: $g$-band optical image from the Fornax Deep Survey \citep[FDS, ][]{Iodice2016, Venhola2018} overlaid with intensity contours of the \hi\ (see Fig. \ref{fig:moment}). Right: far-ultraviolet image from GALEX observations overlaid with velocity contours extracted from the 21 cm observations (Fig. 2). The blue circle indicates what appears to be a large star forming region.}
\label{fig:optical}
\end{figure*}

The Fornax cluster is one of the nearest clusters \citep[D = 20 $\pm$ 0.3 $\pm$ 1.4  Mpc, ][]{Blakeslee2009} which makes it an excellent target to study the effects of the cluster environment on the neutral hydrogen content of its members. This cluster has a high central surface density of galaxies \citep{Ferguson1989} but is rather low mass \citep[${\rm M_{vir}} = 5\times10^{13}$ \msun,][]{Drinkwater2001} possibly forming a bridging environment between massive groups and clusters. The MeerKAT Fornax Survey \citep[MFS, ][]{Serra2023} maps its 21 cm emission to determine the gas distribution and kinematics in this unique environment. As the gas content of cluster galaxies is highly skewed towards gas poor galaxies \citep{Dressler1980, Kleiner2023} it is crucial to establish cluster membership for the relatively few galaxies detected in \hi.\\
\indent The galaxy ESO\,358-60 is an Irregular dwarf galaxy that appears edge-on in the optical \citep[see Fig. \ref{fig:optical},][]{deVaucouleurs1992}. It has a diffuse and fairly uniform surface brightness \citep{Matthews1997} and is star forming at a low rate \citep{Leroy2019}. Table \ref{tab:measures} provides an overview of several of its parameters. It is classified as a likely member of the Fornax cluster \citep[FCC 302,][]{Ferguson1989} following the morphological criteria developed by \cite{Binggeli1985} for determining membership to the Virgo cluster. In the absence of systemic velocities, for irregular galaxies, these criteria are "resolution into knots" of stellar associations and H {\sc ii} regions and adhering to the "surface brightness" versus absolute magnitude relation \citep{Binggeli1984} in the faint end. These criteria are mostly considered for the removal of background galaxies as for Virgo only a small amount or no foreground galaxies are expected \citep{Binggeli1985}. It is not clear how well these criteria exclude foreground galaxies and \cite{Ferguson1989} do not address this when establishing the original membership designations. \\
\indent ESO\,358-60 stands out from other \hi\ detections in the MFS due to its low systemic velocity \citep[${\rm v_{sys}} = 804$ \kms,][]{Schroder2001} and the fact that it does not display signs of tidal interaction or an \hi\ tail  \citep{Serra2023}. Its \hi\ disk is significantly more extended than its observed stellar disk (Fig. \ref{fig:optical}) and it is difficult to reconcile this apparently unaffected disk with the galaxy traveling through the intra-cluster medium (ICM) of the Fornax cluster at a speed of $\ge$ 600 \kms. In fact, if we assume a density of the ICM $\sim 4 \times 10^{-5} {\rm\hspace{0.1cm}cm}^{-3}$ at the projected distance of ESO\,358-60 \citep{Chen2007}, approximate an outer stellar surface brightness $\sim$ 0.4 \msun\ pc$^{-2}$ \citep[$\sim$10\% of the surface brightness at R$_{80}$,][]{Bouquin2018}, and a outer gas surface brightness of 0.4  \msun\ pc$^{-2}$ (this work), following the prescription from \cite{Gunn1972}, we find a ratio of ram pressure to restoring force $\sim$ 100. This ratio is far too high for the gas disk of the galaxy to remain unaffected by the ram pressure.\\
\indent One possible explanation for these peculiarities would be that ESO\,358-60 is not part of the Fornax cluster. Its low systemic velocity does not exclude it immediately as a Fornax member as its difference with the systemic velocity of the cluster center \citep[$v_{\rm sys}$ = 1454 \kms,][]{Maddox2019}  is slightly more than two times the velocity dispersion of the cluster \citep[$\sigma_{\rm Fornax} = 286$ \kms,][]{Maddox2019}. Therefore, taking advantage of the extremely regular morphology and kinematics of the \hi\ disk (Sect. \ref{Data}), we perform an in depth investigation of the neutral hydrogen in this galaxy in order to place it on the baryonic Tully-Fisher relation \citep[BTFR, ][]{Tully1977, McGaugh2000, Lelli2019}. Due to its low scatter, the BTFR can be used to obtain the baryonic mass of the galaxy based solely on its rotational velocity. This mass can then be equated to the mass obtained from the flux of the stellar and gas components to calculate a distance independent of redshift \citep{Tully1977}.\\
\indent This paper is structured as follows; In $\S$ \ref{Data} we present and discuss the \hi\ data that we use to obtain a tilted ring model (TRM) and investigate the surroundings of ESO\,358-60, $\S$ \ref{Models} explains how we obtained the TRM. In $\S$ \ref{Distance} we derive a distance from the BTFR and compare it to other distance indicators, in $\S$ \ref{RadialMotions} we present and discuss the presence of radial motions in this galaxy and finally in $\S$ \ref{Conclusions} we summarize our findings.\\
\section{Data}\label{Data}
\begin{figure*}[htbp!]
\hspace*{-1.5cm}\includegraphics[width=2.25\columnwidth]{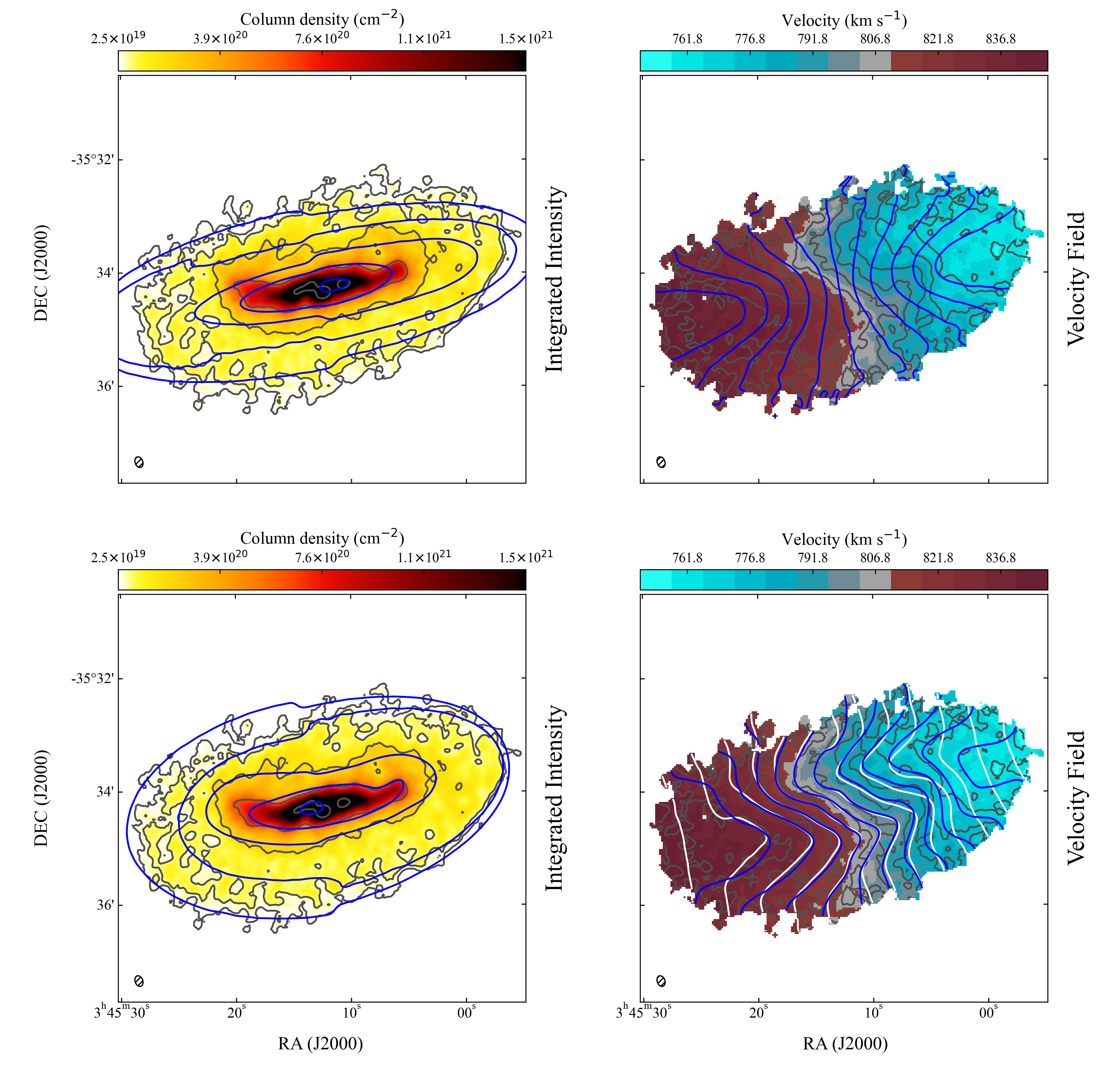}
\caption{Moment maps of the \hi\ in the galaxy ESO\,358-60.  Dark grey contours indicate the data and the blue contours the automatic model in the top panels and in bottom panels the best fit model. Left: moment 0 or integrated intensity map  of the galaxy, contours are at $\sqrt{10^{n}}$ $ \times$ 2 $
 \times$ $10^{19} {\rm cm}^{-2}$. Right: moment 1 or velocity field, contours are centered on the systemic velocity (V$_{\rm sys}$ = 806.8 \kms) and decrease (increase) in steps of 7.5 \kms\ for the approaching (receding) side. The white contours represent the best fit model with a rising rotation curve (see text). The ellipse in the bottom left corner of the panels indicates the restoring beam of the observations.}
 \label{fig:moment}
\end{figure*}
\begin{figure}[htbp!]
\includegraphics[width=1\columnwidth]{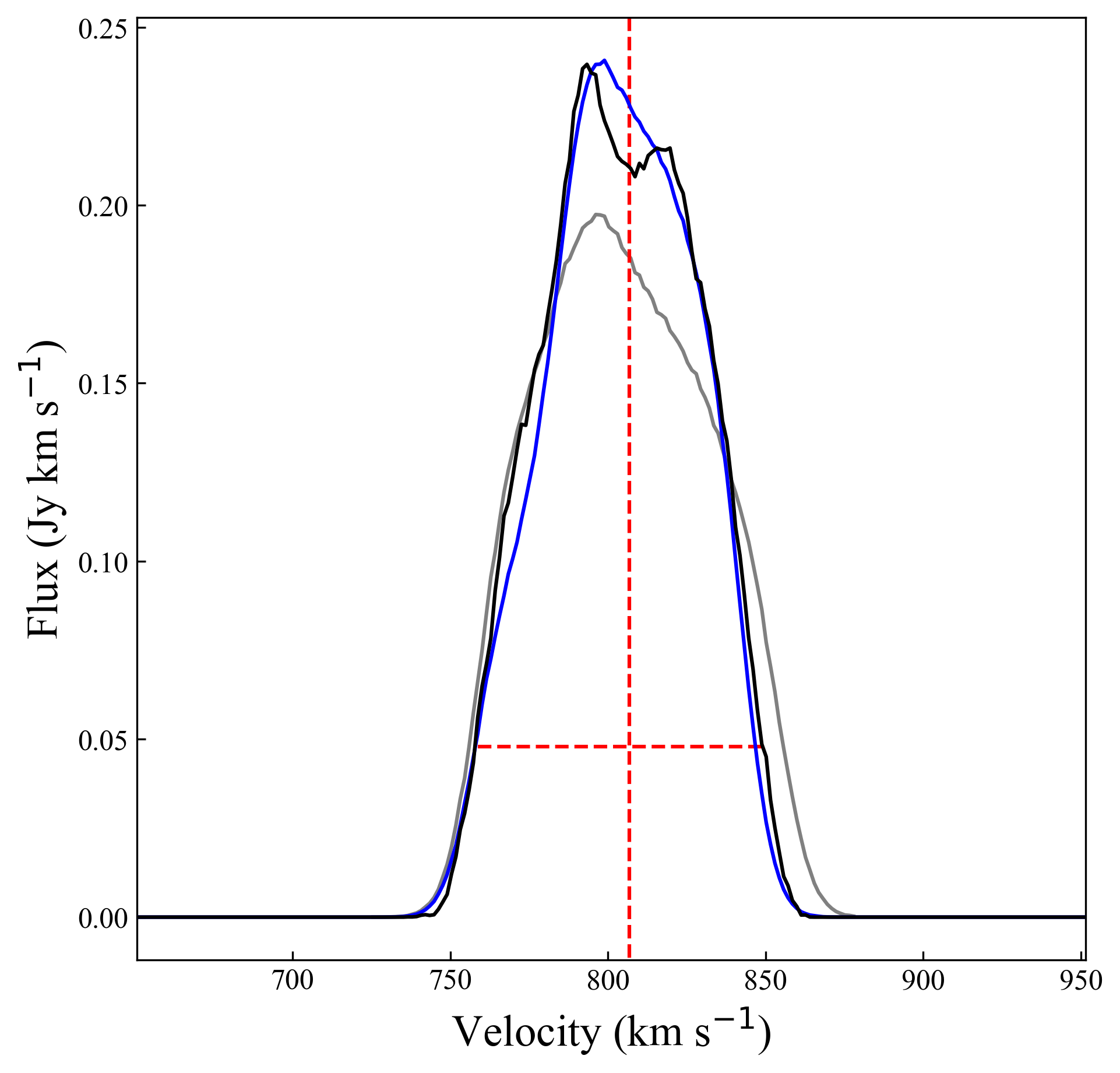}
\caption{Global line profile of ESO\,358-60. The black line indicates the data, the blue the best fit model and the grey the automatic model. The vertical red dashed line indicates the systemic velocity, and the horizontal line the uncorrected \wfif.}
 \label{fig:global}
\end{figure}
\begin{figure*}[htbp!]
\includegraphics[width=2.\columnwidth]{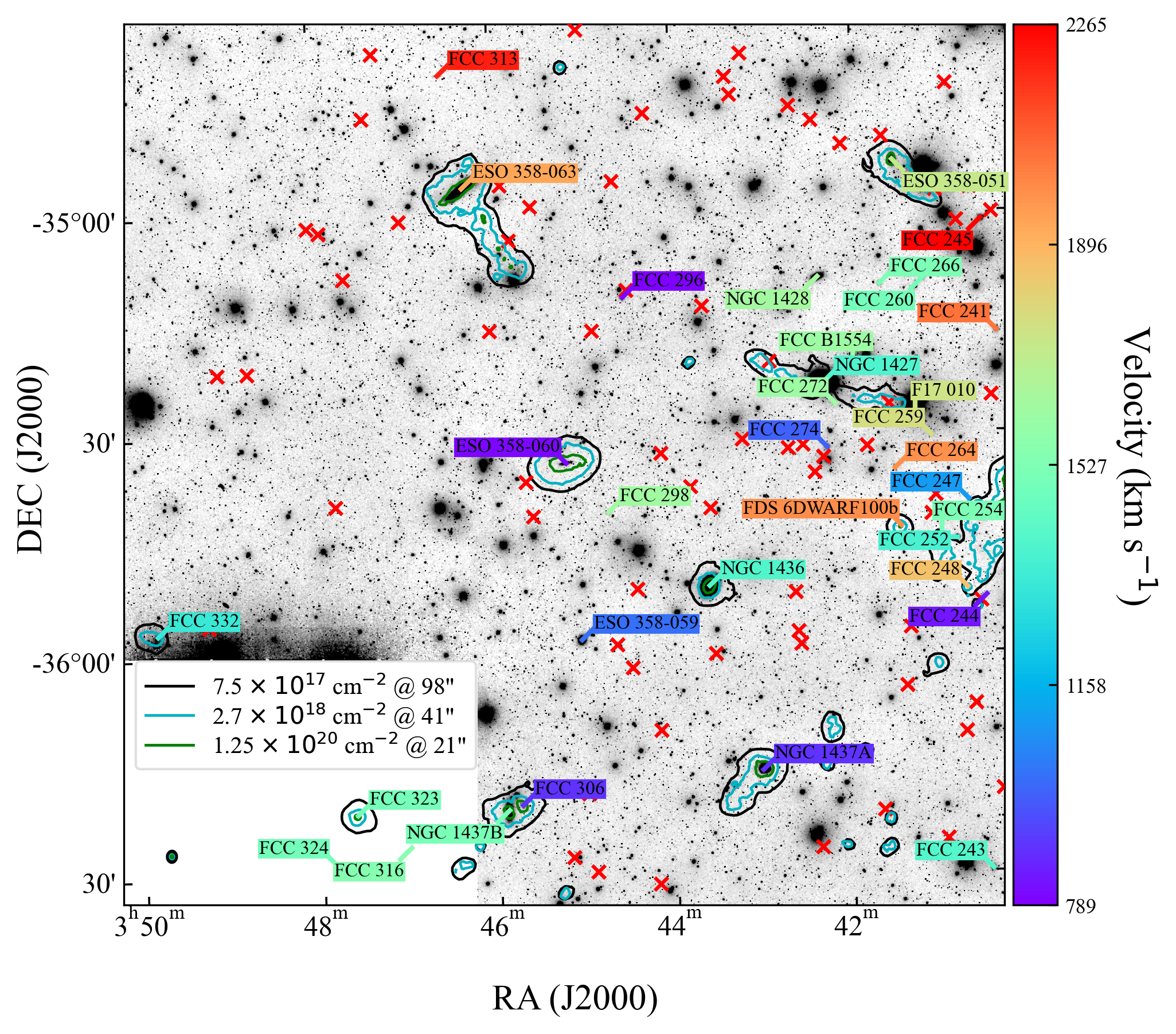}
\caption{$g'$-band image of the 2 deg$^2$ surrounding ESO\,358-60 from the FDS \citep{Iodice2016,Venhola2018} overlaid with the lowest reliable contours (3$\sigma$) of the 98" and 41" resolutions  of the MFS.  For the 21" resolution we plot the contour corresponding to $\sim$ 1 \msun\ pc$^{-2}$. The beams are displayed in the bottom left corner. All galaxies that are spectroscopically confirmed in the field are labeled, where the background color of the label indicates the systemic velocity of the system. The color bar on the right shows the correspondence between color and velocities. Galaxies identified in the FDS, but not spectroscopically confirmed, are marked with a red x. }
 \label{fig:overview}
\end{figure*}

\begin{table}
    \centering
    \begin{tabular}{@{} lll @{}} 
       \multicolumn{3}{c}{ESO\,358-60}\\
       \hline
       \hline
	Parameter&  & Ref.\\
       \hline
       	Morphological Type & Irregular & 2\\
	Center ($\alpha$ J2000)&$3^{{\rm h}}$  45$^{{\rm m}}$$13^{{\rm s}}$& 1\\
	 \hspace*{0.99cm}($\delta$ J2000)&$-35^{\circ}$34\arcmin\ 17\arcsec & 1\\
	v$_{\rm sys}$ (km s$^{-1})$& 806.8 $\pm$ 0.8& 1\\              		
    m$_{r'}$ (mag) & 15.5 $\pm$ 0.2 & 5\\
    m$_{3.6\mu m}$ (mag) & 16.549 $\pm$ 0.004 & 3\\
    $g'-r'$ (mag) & 0.17 & 5\\
    SFR (\msun\ yr$^-1$)$^{\rm{a}}$&0.007 $\pm$ 0.004 &4 \\
	Optical PA (\degr)& 102 &2\\
    R$_{{\rm eff} r'}$ (arcsec) & 30.3 $\pm$ 6.3 & 5   \\
	D$_{25}$ (arcsec)& 104.3&2\\
	D$_{\rm{H \textsc{i}}}$ (arcsec)$^{\rm{b}}$&221 &1\\
	D$_{\rm{H \textsc{i}}}$/D$_{25}$&2.1&1\\
	F$_{\hi}$ (Jy \kms) &   12.0 $\pm$ 1.2 & 1 \\
	V$_{\rm max}$ (km s$^{-1}$)& 48.1 $\pm$ 2.0 & 1\\
	\vflat\ (km s$^{-1}$)$^{\rm c}$& 48.1 $\pm$ 1.4 & 1\\
    V$_{\HI}$ (km s$^{-1}$)& 34.8 $\pm$ 2.1 & 1\\
    \wfif\ (km s$^{-1}$)& 91.8 $\pm$ 1.4 & 1\\
 \hline
    \end{tabular}
     \caption{Properties of ESO\,358-60. The numbers in the column of references refer to: (1) this work, (2) \cite{deVaucouleurs1992}, (3) \cite{Bouquin2018}, (4) \cite{Leroy2019}, (5) \cite{Venhola2018}. a) Re-scaled to our final best fit distance of 9.4 Mpc. b) The diameter of the galaxy at a column density of 1 \msun\ pc$^{-2}$ from the best fit model. c) The error is taken as the mean of all rings with \vflat. }
\label{tab:measures}
\end{table}
The \hi\ data used in this study is observed as part of the MFS. For the tilted ring modeling we focus on the highest resolution data cube for ESO\,358-60. We use the lower resolution data products solely to obtain an impression of the \hi\ environment of the galaxy at the end of this section. For the details of the data reduction and the lower resolution product we refer to \cite{Serra2023}.\\
\indent For ESO 358-60 the high resolution data cube has a spatial resolution of 12.2" $\times$ 9.6" FWHM, a velocity resolution of 1.4 \kms\  and a noise level of 0.24 mJy beam$^{-1}$ in a channel (n$_{\hi}$ = 4 $\times$ $10^{19}$ cm$^{-2}$ for 3$\sigma$ over a 25 \kms\ line width). This is lower than the average noise across the full mosaic \citep[$\sigma_{\rm full}=$0.3 mJy beam$^{-1}$,][]{Serra2023} but within the expected variations of the noise level across the MFS footprint. The cube is in a barycentric spectroscopic frame and all velocities cited in this paper follow the optical definition.\\
\indent Fig. \ref{fig:moment} shows the moment 0 and moment 1 map of the data cube and Fig. \ref{fig:global} the global line profile. Already by visually inspecting these maps several obvious implications can be drawn for the distribution and kinematics of the \hi\ in this galaxy. The central surface brightness distribution is highly elongated indicating an edge-on disk or a bar like distribution in the center. The outer parts are much more circular and thus point to a lower inclination. The surface brightness distribution in the approaching and receding sides appear largely symmetric and, except for  a slight position angle (PA) warp in the receding side, no major deviations from an elliptical distribution are visible. The peak of the \hi\ emission coincides with a large star forming region, indicated by the blue circle in Fig. \ref{fig:optical}, just offset from the center of the galaxy. The galaxy does look somewhat disturbed in the optical but this is not unusual for an irregular galaxy. The most prominent disturbance lies on the Eastern edge on the optical disk; just before the start of the PA warp in the receding side of the galaxy.\\
\indent The velocity field shows a similar picture of regularity. The central major and minor axes are not perpendicular indicating the existence of radial motions in the galaxy. These motions appear to occur largely symmetrically around the morphological minor axis and cover the full extent of the optical disk. Besides these radial motions the velocity field appears regular and no major disturbances can be identified in this map nor in the cube directly.\\ 
\indent We use the lower resolution products of the MFS to search for \hi\ that could indicate an interaction despite the regularity of the disk. Fig. \ref{fig:overview} shows the 2 deg$^2$ surrounding ESO\,358-60 and shows the relatively quiet immediate surroundings of the galaxy compared to other Fornax galaxies detected in \hi. In fact the only spectroscopically confirmed galaxy within $\sim$ 20' ($\sim$ 115 kpc at the distance of Fornax) is FCC 298 which is significantly offset in velocity (V$_{\rm sys}$=1620 \kms).  Within a 20' radius  there are six more dwarfs for which no redshift is observed \citep{Venhola2018}. These are all significantly redder ($g'-r'$ > 0.5) and fainter (m$_{r'}$ > 17) than ESO\,358-60 and their effective radius in the $r'$- band is always less than half (R$_{{\rm eff} r'}$ < 14") that of ESO\,358-60 (see Table \ref{tab:measures}). None of these dwarfs are detected in the MFS nor have they visibly affected the \hi\ disk of ESO\,358-60. The nearest galaxy detected in \hi\ is NGC\,1436, a lenticular galaxy for which the outer \hi\ disk was stripped and the detected \hi\ is well contained within the stellar disk \citep{Loni2023}. All other galaxies in Fig. \ref{fig:overview} detected in \hi\ show disturbed disks and tails and \hi\ debris. It is these differences with its surrounding \hi\ detections that have prompted us to question the cluster membership of ESO\,358-60.\\     
\indent As mentioned in the previous paragraph, in the cluster environment it is still possible to have a fairly regular disk that has had its outer edges stripped away by interactions \citep{Chung2009, Loni2023}. Therefore, as final check , we compare the \hi\ diameter to the optical diameter. The ratio between these two diameters is 2.1, a ratio that is normal for field galaxies \citep[Table \ref{tab:measures}, ][]{Broeils1997}. This confirms that the \hi\ of ESO\,358-60 does not show significant distortions when compared to field galaxies.\\
\section{Models}\label{Models}
\begin{figure}[htbp!]
\includegraphics[width=0.99\columnwidth]{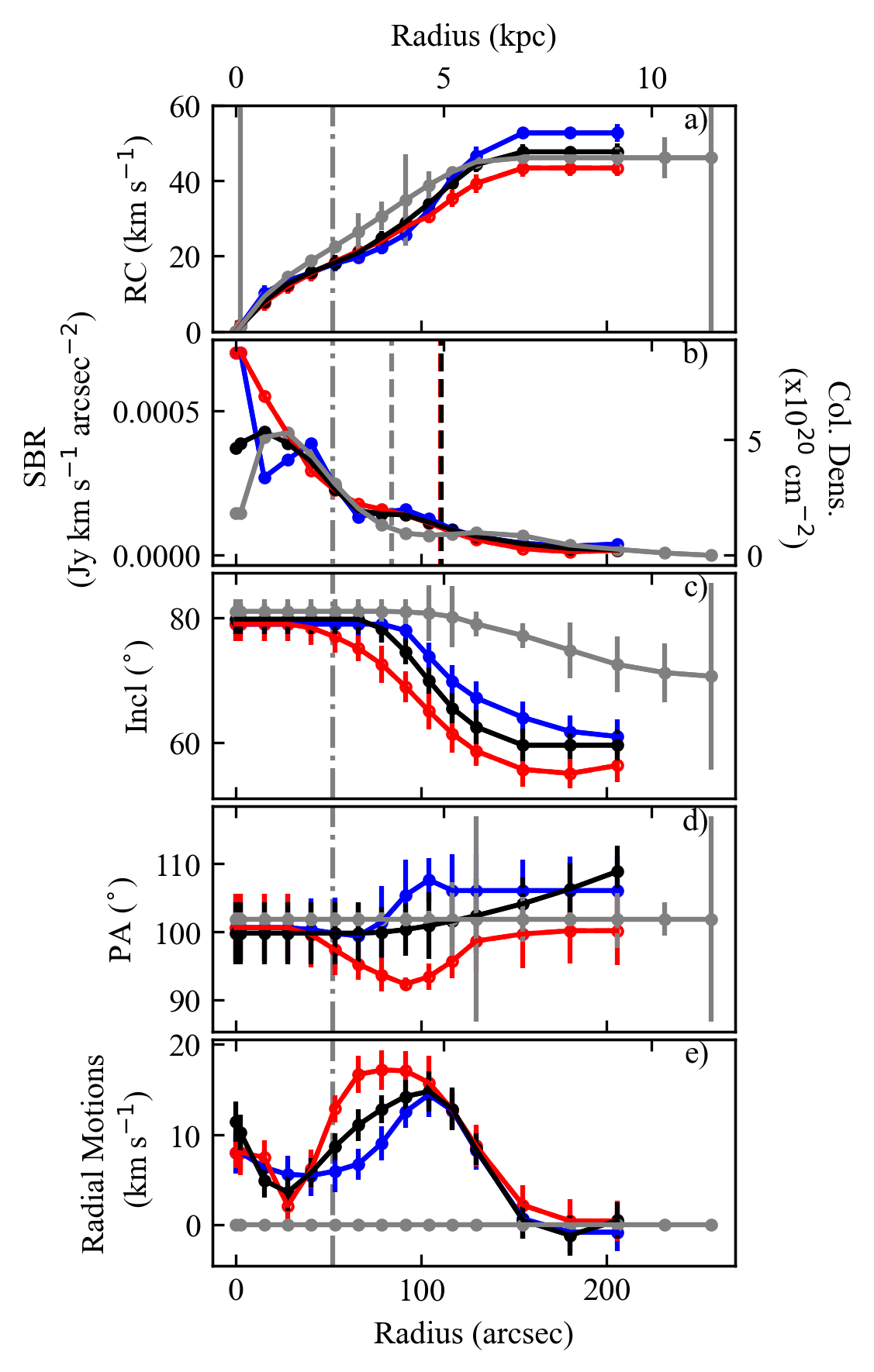}   
\caption{The fitted parameters for the various models. Light gray lines represent the {\sc pyFAT} model. Black lines the single disk model and the blue (approaching) and red (receding) lines the final best fit model. Panel a) is the rotation velocities, b) surface brightness distribution, c) inclination, d) PA, e) radial motions, all as function of radius. The vertical gray dot dashed line indicates R$_{25}$ and the vertical dashed lines in panel b) indicate where the surface brightness is equal to 1 \msun\ pc$^{-2}$.} 
\label{fig:parameters}
\end{figure}

\begin{figure}[htbp!]
\includegraphics[width=0.99\columnwidth]{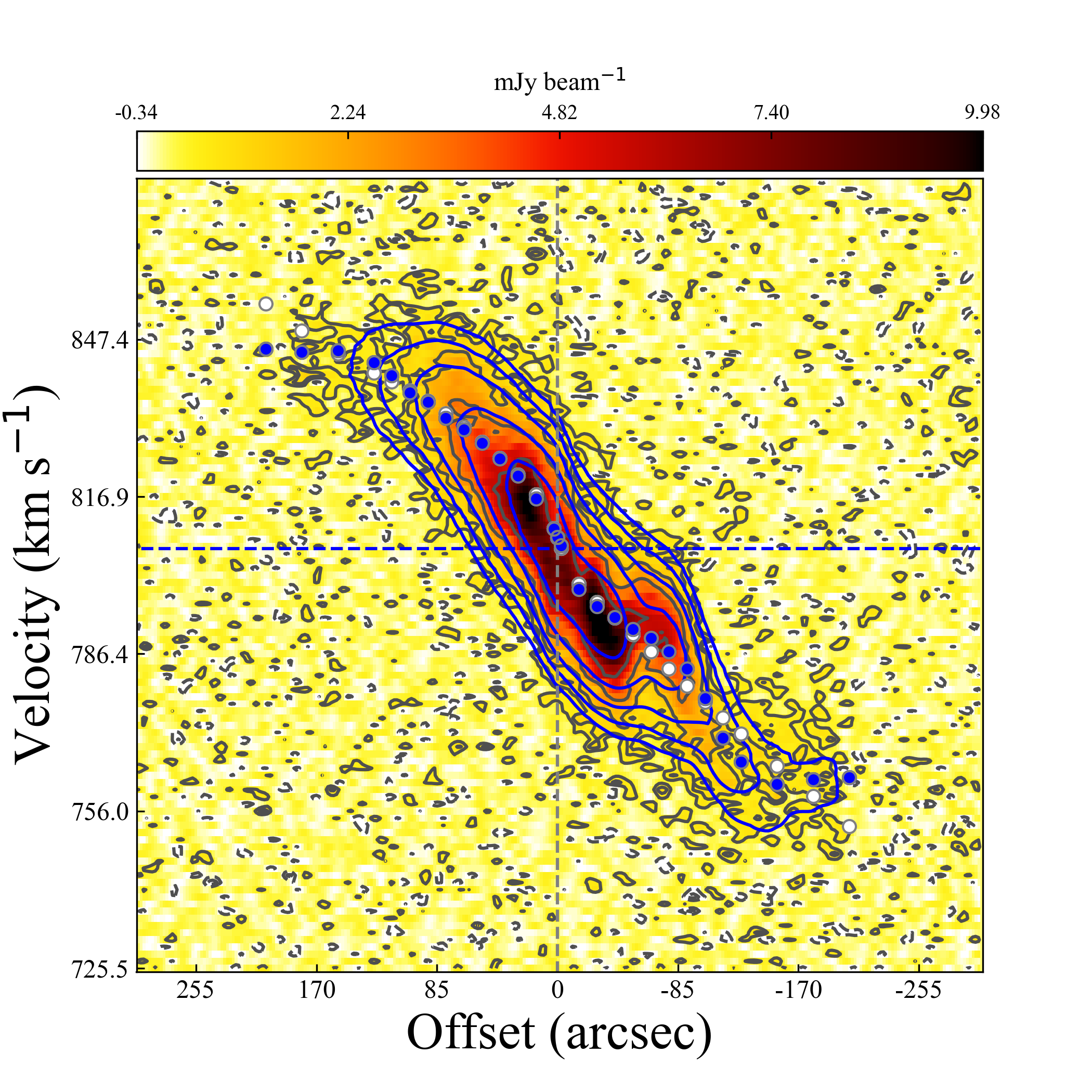}   
\caption{PV-diagram along the major axis. Gray contours and the color scale show the data, blue contours show the best fit model. Contours are at 1.5, 3, 6, 12, 24 $\times$ $\sigma$ with $\sigma$ = 0.24 mJy beam$^{-1}$ (3.2 $\times 10^{18}\ {\rm cm}^{-2}$ in a channel). The gray dashed line indicates the spatial fitted center and the colored dashed line indicates the fitted systemic velocity of the best fit model. The blue circles indicate the V$_{{\rm obs}}={\rm V_{rot}\times\ sin}(i)$ of the best fit model and the white circles indicate the same for the rising model (see text). To be concurrent with the right ascension, offsets are decreasing from left to right.}   \label{fig:pvdiagram}
\end{figure}

In order to obtain an initial TRM for the galaxy we first run {\sc pyFAT} \citep[v0.1.2\footnote{\url{https://github.com/PeterKamphuis/pyFAT-astro}},][]{Kamphuis2015,Kamphuis2024} on the data cube. The parameters of this automatic fit are shown in Fig. \ref{fig:parameters} as the gray lines and the non-radially varying parameters are listed in Table \ref{tab:parameters}. Even though {\sc pyFAT} divides the model in an approaching and receding side, the final preferred model is fully symmetric, reflecting the regularity and symmetry that made ESO\,358-60 stand out in the first place. The moment 0 and 1 maps of the automated model are displayed as the blue contours in the top panels of Fig. \ref{fig:moment}. This model fits the data well within the parameters that are allowed to vary within {\sc pyFAT} with a few  exceptions. In the top left panel of Fig. \ref{fig:moment} it can be seen that the model is too elongated compared to the data. This is due to a surface brightness that is too high in the outer rings and, even though a significant variation of the inclination ($\Delta i\ \sim 10^\circ$) is already incorporated in the model, an underestimation of the line of sight warp. Additionally, the radial motions are not fitted by {\sc pyFAT} (see Fig. \ref{fig:moment}, top right) and need to be incorporated into the model manually. We therefore consider this model as a starting point,  and improve on it through guided fitting with \tirific\ \citep{Jozsa2007}.\\
\indent As the automatic fit is completely symmetric we first reduce the model back to a single disk. After manually increasing the line of sight warping in the input parameters we refit the model to the data. During the manual fitting we group certain rings for several parameters, through \tirific's VARINDX parameter, to ensure a stable fit. The grouping varies between different parameters in order to reduce degeneracy within the fitting. For example, in the inclination we might group the rings 10 to 13 by only varying rings 10 and 13 and determining rings 11 and 12 through a non-rounded Akima interpolation with natural boundaries. In this case we would make sure that in the rotational velocities we fit either ring 11 or 12 and interpolate over rings 10 and 13 in order to counteract the degeneracy between these parameters in the fitting. If the outer most rings are not varied individually we extrapolate by matching the ring value to the last fitted ring. We always run \tirific\ in a fitting mode and judge the quality of the model by a visual comparison between various representations of the data and the model. Between iterations we might manually adjust the limits in the input parameter file but any final model is always the result of the $\chi^2$ minimization process with \tiri. \\
\begin{table}
    \centering
    \begin{tabular}{@{} llll @{}} 
       \multicolumn{4}{c}{Non-radially varying model parameters}\\
       \hline
       \hline
	Parameter&{\sc pyFAT}&Single disk& Best Fit\\
       \hline
Center ($\alpha$ J2000)&$3^{{\rm h}}$  45$^{{\rm m}}$$13^{{\rm s}}$& 
$3^{{\rm h}}$  45$^{{\rm m}}$$13^{{\rm s}}$
&
$3^{{\rm h}}$  45$^{{\rm m}}$$13^{{\rm s}}$
\\
\hspace*{0.99cm}($\delta$ J2000)&$-35^{\circ}$34\arcmin\ 17\arcsec & $-35^{\circ}$34\arcmin\ 17\arcsec
& $-35^{\circ}$34\arcmin\ 17\arcsec\\
	v$_{\rm sys}$ (km s$^{-1})$& 806.9 $\pm$ 0.8& 805.8 $\pm$ 0.9 & 806.8 $\pm$ 0.8\\   
       Scale height (\arcsec)&8.6 $\pm$ 1.2&8.6  &  8.6 \\
      $\sigma$ (\kms)$^{a}$&  10.3-8.6 &8.0 $\pm$ 2.2&8.0 $\pm$ 1.6\\
        \\
 \hline
    \end{tabular}
     \caption{Parameters that are not varied as function of radius for the manual fitting. $a)$ In the automated fit the dispersion of the model can vary as function of radius but as we fit it as a singular parameter in the manual fitting it is included in this table.  }
     \label{tab:parameters}
 \end{table}
\indent Compared to the fitting philosophy of {\sc pyFAT} we make two more changes to ensure the stability of the fitting while we add radial motions to the rings. Firstly, we fit the dispersion as a singular parameter. Secondly, we do not allow the scale height of the model to vary compared to the value found in the automatic fit.\\ 
\indent After a few iterations it becomes clear that the required surface brightness of the outer rings should be so far below the noise of the data that they have no influence on the model anymore. Therefore we cut the last two rings from the modeling. Once we deem the outer warp to be well matched in the model we add radial motions to the model.\\
\indent This procedure leads to a single disk best fit model of which the parameters are shown as the black lines in  Fig. \ref{fig:parameters}. However, after including the radial motions, it turns out that the \hi\ distribution in the galaxy is not perfectly symmetric and it looks like the model can be improved by fitting the approaching and receding side independently. As such we split the model into two disks again and refit this model  in the same manner as the manual fits for the single disk. This leads to a final best fit model of which the parameters are shown in Fig. \ref{fig:parameters} as the blue (approaching) and red (receding) lines. The bottom panels of Fig. \ref{fig:moment} show the model overlaid on the moment maps and Fig.  \ref{fig:pvdiagram} a Position-Velocity (PV) diagram along the major axis. \\
\indent This model fits the data remarkably well as can be seen in the moment maps (Fig. \ref{fig:moment}, bottom panels, blue contours) and is our best fit model. As in several smaller galaxies rotation curves (RCs) are observed to keep rising until their last measured point we refit our best fit model with \tiri\ forcing a RC that continues to rise until the last point.\\
\indent Visually comparing this model to the data shows that such a model deteriorates the match between the data and model in the outer rings. This is most clearly seen in the velocity field (Fig. \ref{fig:moment}, right bottom panel, white contours) where the outer contours are more open than the data. As such we consider the RCs of the rising model as an absolute upper limit on the RC.\\
\indent In the analysis that follows we use the model with a flat outer RC where the approaching and receding side are fitted independently of each other as this is the best fit model. However, as the single disk model is mostly an average between the two sides (see Fig. \ref{fig:parameters}) it makes little difference for the analysis of the galaxy as whole whether we use our best fit model or the single disk model. \\
\indent In the manual fitting we fit the dispersion as a singular value and have fixed the scale height.  We do not expect the dispersion to affect the fitted rotational velocities but the scale height could.  The scale height we use (see Table \ref{tab:parameters}) is $\sim$ 0.4 kpc around 10 Mpc. 
This is typical for the inner parts of \hi\ disks \citep{O'Brien2010}\footnote{We use a $sech^2$ distribution in the model which means the FWHM is $\sim 1.7627\times {z_0} = 0.7$ kpc.}. However, \hi\ disks are expected to flare significantly. If this is the case in ESO 358-60 we might underestimate the scale height in the outer parts of the model. In principle this can then lead to us underestimating the outer inclinations as the increased thickness of the disk is incorrectly interpreted as an increased circularity and thus a lower inclination. If the true outer inclinations are actually higher than in our best fit model the rotational velocities in the outer parts, and thus \vflat\ would be lower.   \\
\indent It is unlikely that  the effect described above significantly affect our results for the following reasons. First, the fit is not only performed on the intensity distribution but also on the kinematics. As long as the gas at increased vertical heights is co-rotating with the gas in the disk, the kinematics should be minimally affected by the flaring. Secondly, as the scale height is fitted in arcseconds, the closer the galaxy is to Fornax the thicker the physical disk in the model is, thus reducing the effect of us not fitting a flare. Finally, if we fully interpret the observed decrease of the isophotal ellipticity in terms of an increase in scale-height and completely remove the change in inclination in the model this would lead to rotational velocities $\sim$ 10$\%$ lower. Such an extreme change in model geometry is beyond what is reasonable to expect, as the warp is already visually identifiable in the data (see Fig. \ref{fig:moment}), but even in this case we would not expect our results to change significantly as other errors dominate the determination of the distance. Additionally, as we will show in the next section, lower rotational velocities lead to a smaller distance as the baryonic mass derived from the BTFR would decline, strengthening our conclusion on the position of ESO 358-60 relative to Fornax along the line of sight. \\
\indent After determining the best fit models we estimate the errors on the models using the python package TRM\textunderscore errors\footnote{v0.0.6, \url{https://github.com/PeterKamphuis/TRM_errors}}. This package runs \tiri\ a given amount of times, in our case 30, with the start values varied with a binomial distribution with three times the width of their initial step size. It then calculates the errors from spread in the best-fitting output. This method provides us with the best possible errors but they do not cover possible conceptual errors, such as a fixed scale height, in our fitting setup.\\
\section{The Distance to ESO\,358-60}\label{Distance}
\begin{figure}[htbp!]
\includegraphics[width=0.99\columnwidth]{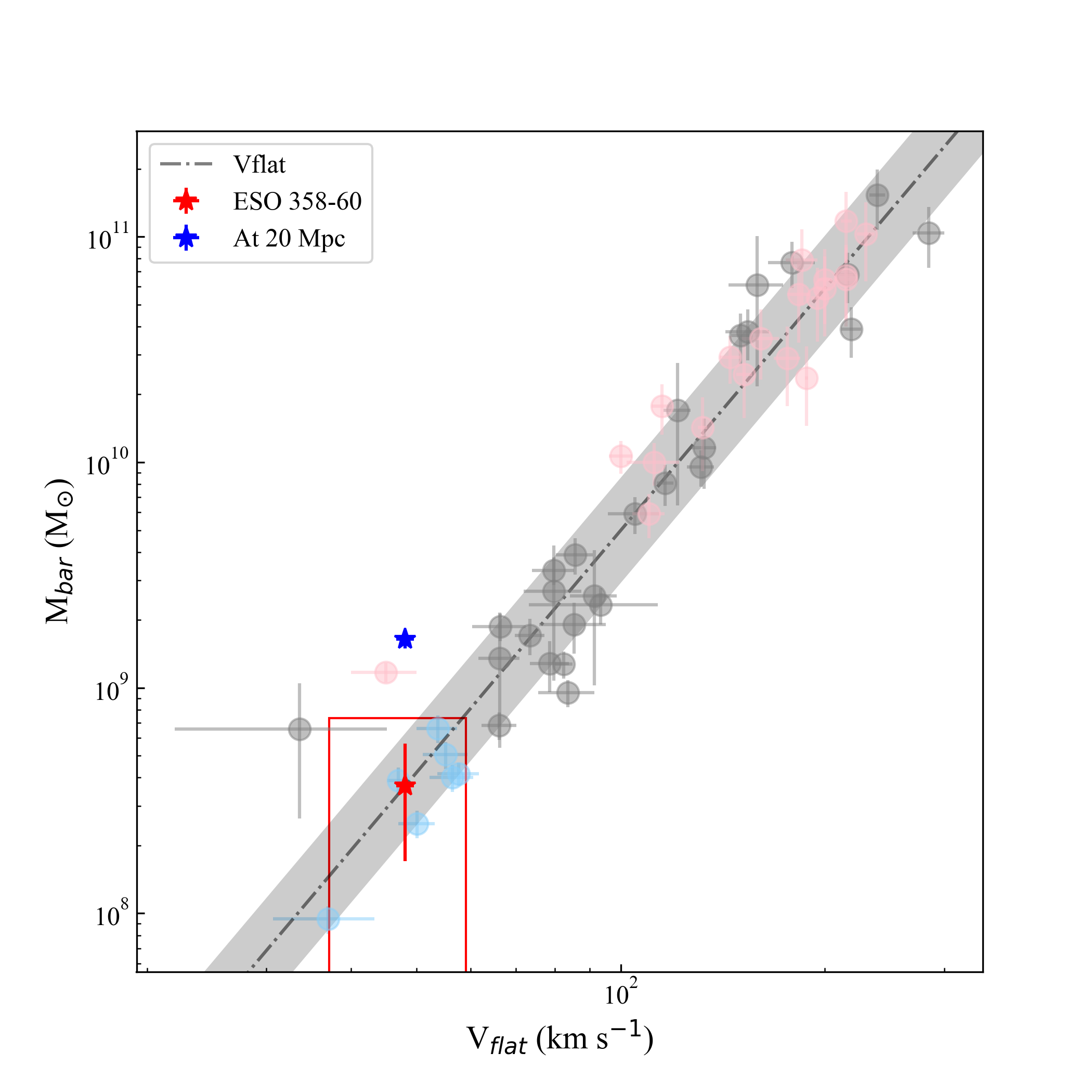}   
\caption{BTFR for the sample of galaxies with redshift independent distances from SPARC \citep[gray+blue points,][]{Lelli2016, Lelli2019} supplemented with galaxies from \cite[][red points]{Ponomareva2016, Ponomareva2018}. The dashed line indicates the relation fitted for \vflat\ with the shaded gray area representing the observed vertical scatter. The stars indicate ESO\,358-60 on the relation for the best fit model (red), at a distance of 20$\pm$1 Mpc (blue). The red box is determined by its top right corner where ESO\,358-60 is positioned on the relation when using the upper limit derived from the rising RC model.}
\label{fig:TF}
\end{figure}

\begin{figure}[tbp!]
\includegraphics[width=0.99\columnwidth]{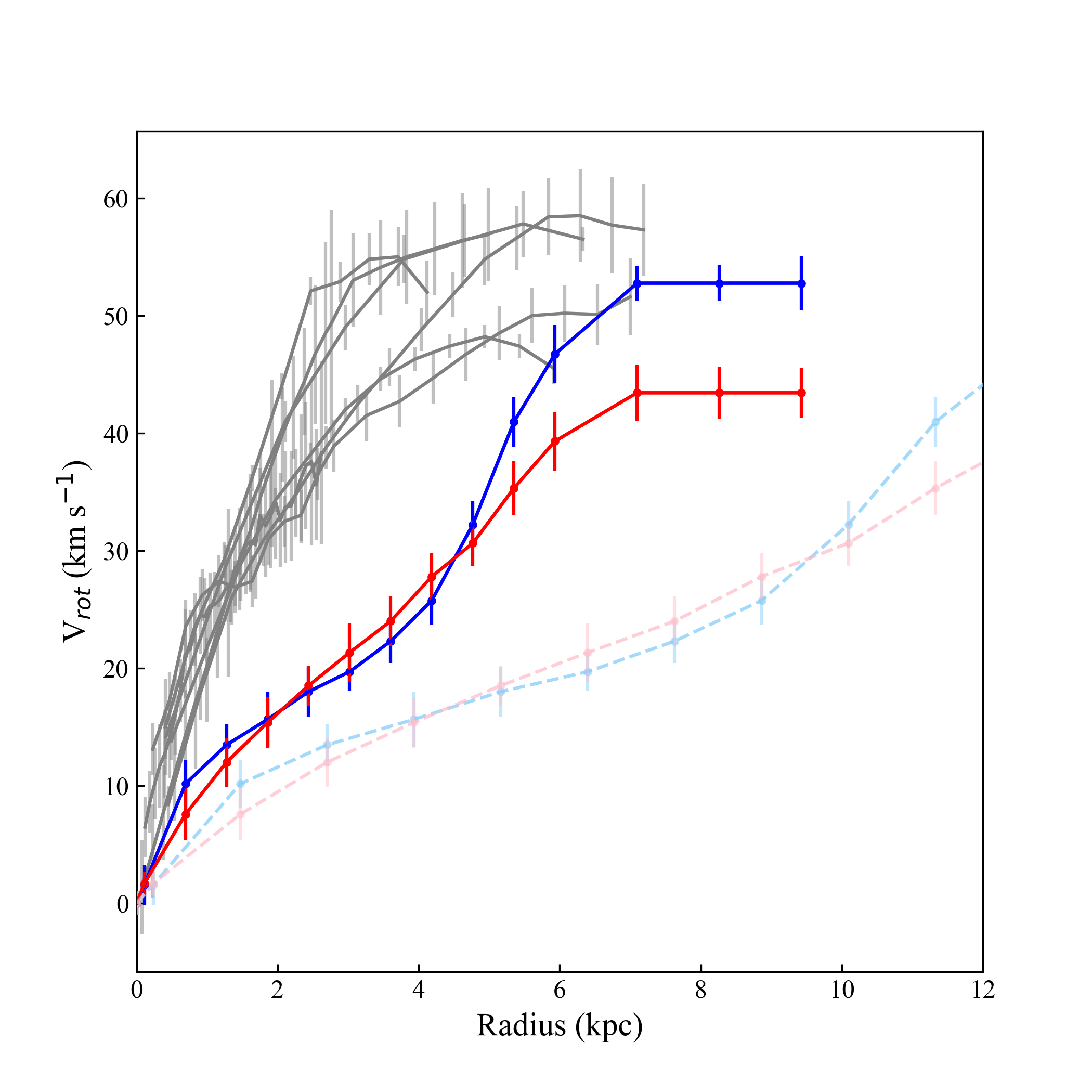}   
\caption{RCs (gray lines) of the SPARC database in the neighborhood of ESO\,358-60 on the TF-Relation. These galaxies are indicated as the light blue points in Fig. \ref{fig:TF}. Blue and red line, the best fit RC of ESO\,358-60 (respectively the approaching and receding sides) at BTFR calculated distance. Light blue and pink lines, the same but now at a distance of 20 Mpc. These continue outside the plotted area.}
\label{fig:RCs}
\end{figure}

\begin{figure}[htbp!]
\includegraphics[width=1.1\columnwidth]{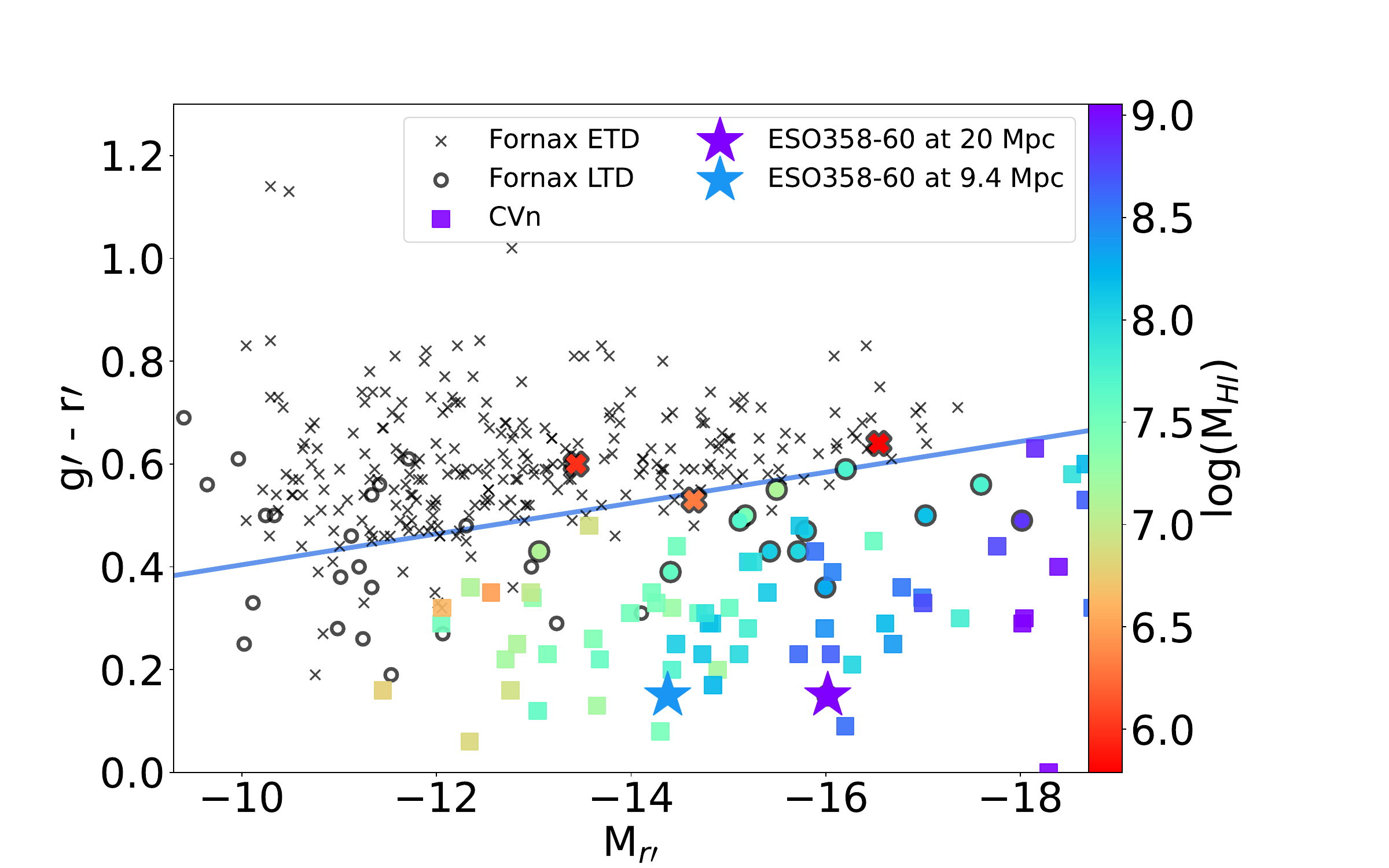} 
\caption{Color-magnitude diagram with the dwarfs in the central 2.5 $\times$ 4 deg$^2$ of the Fornax cluster (crosses indicate early type dwarfs and circles late type dwarfs) and those in the Canes Venatici region (squares). For the Fornax cluster filled colored symbols are dwarfs with \hi\ whereas  the black symbols are undetected in \hi. The star symbols indicate ESO\,358-60 at the best fit distance (blue) and at 20 Mpc (purple). This figure is an updated version of Fig. 4. in \cite{Kleiner2023} with the Canes Venatici sample added and ESO\,358-60 highlighted. }
\label{fig:CM-diagram}
\end{figure}

In order to determine the distance to ESO\,358-60 we follow a procedure similar to the one presented in \cite{Schombert2020}. We start with the data presented in the Spitzer Photometry and Accurate Rotation Curves (SPARC) database \citep{Lelli2016, Lelli2019}.  The reason for using SPARC is threefold, a) it uses resolved rotation curves, b) it samples the low RC domain ($\le 40$ \kms) better than other resolved studies \citep[e.g.][]{Ponomareva2018}, c) all input to this BTFR is available online\footnote{http://astroweb.cwru.edu/SPARC/} allowing for an in depth comparison between the RC of ESO\,358-60 and those used to construct the relationship. Unfortunately, many of the distances presented in SPARC are not redshift independent distances. Therefore we cross-correlate the SPARC database  with the NED-D database \citep[v17.1.0,][]{Steer2017} and only retain those galaxies that have a distance determined through the tip of the red giant branch or cepheids. If a galaxy has multiple distance measurements based on these methods we take the mean of these as the distance to said galaxy. We re-scale the distance dependent values of the SPARC database to these new distances. This results in 46 SPARC galaxies with redshift independent distances. In order to increase the sample size, we add to our sample the galaxies from \cite[][all with distances from the tip of the red giant branch or cepheids]{Ponomareva2016, Ponomareva2018} that are not already present in the SPARC database. This leads to an additional 21 galaxies. From these 67 galaxies we determine a BTFR specifically intended for the derivation of a baryonic mass.\\
\indent The BTFR can be constructed with different indicators for the circular velocity which results in different slopes and scatter for said indicators \citep{Lelli2019, Ponomareva2018}. Additionally, different fitting routines and the inclusion of intrinsic scatter can lead to different results as well  \citep{Bradford2016, Lelli2019}. Therefore, BTFRs cannot be simply compared through the various power law fits available in the literature. However, the baryonic mass derived from the galaxy's circular velocity through the BTFR is accurate as long the input data, for galaxy and BTFR, are consistent. For these reasons we re-derive the BTFR, defined in our dataset, through the {\sc BayesLineFit} software \citep{Lelli2019}. \\
\begin{table}
    \centering
    
    \begin{tabular}{@{} lllll @{}} 
       \multicolumn{5}{c}{BTFR parameters}\\
       \hline
       \hline
       &\vflat&\vmax&V$_{\hi}$&\wfif\\
       \hline
Slope&3.6$\pm$0.1&3.2$\pm$0.1&2.6$\pm$0.2&3.7$\pm$0.2 \\ 
Intercept&2.6$\pm$0.3&3.2$\pm$0.2&4.6$\pm$0.5&0.9$\pm$0.4 \\ 
Obs. vert. scatter&0.23&0.30&0.36&0.38 \\ 
No. galaxies&54&67&61&66 \\ 
                        
 \hline
    \end{tabular}
     \caption{Slope, intercept, vertical scatter and number of galaxies for each fit to the combined sample. }
     \label{tab:btfrfit}
 \end{table}
\indent To obtain a BTFR that delivers the most accurate baryonic mass estimate for ESO\,356-60 it is important to minimize the vertical distance of the data. For this purpose we only add vertical intrinsic scatter to the fit and we refit the BTFR for \vmax, \wfif\ and \vflat. We use these values because i)\vflat\ is found to provide the tightest relation for an orthogonal fit to the SPARC database \citep{Lelli2019}, and ii) \vmax\ is the resolved parameter available to all galaxies in the combined sample. We also investigate \wfif\, corrected for inclination, as it is a parameter used in many Tully-Fisher (TF) relation studies and \cite{Ponomareva2018} find W$_{50}$ to have the tightest relation\footnote{We use \wfif\ as W$_{50}$ is not available in SPARC. Wm$_{50}$ is available in SPARC but it is based on the mean flux of the profile not the peak flux.}  We also compare with the recently published relation for V$_{\hi}$ for the Widefield ASKAP L-band Legacy All-sky Blind surveY (WALLABY) pilot phase observations as recently presented in \cite{Deg2024}.\\
\indent The results for the different fits are presented in Table \ref{tab:btfrfit}. We find that for our sample the observed vertical scatter for the different fits is the smallest for \vflat. Using this relation we find M$_{\rm bar}$ =  4 $\pm$ 2 $\times$  $10^8$ M$_{\odot}$, where the error is calculated from the observed vertical scatter on the BTFR. Fig. \ref{fig:TF} shows the location of ESO\,358-60 as a red star on top of the BTFR for \vflat. The other BTFRs investigated here provide similar results for the derived baryonic mass, albeit with larger errors due to the increased scatter.  \\
\indent To derive a distance from this baryonic mass we need the individual components that contribute to it, i.e. the stellar and gas mass. Using the total flux from the \hi\ integrated intensity map (Fig. \ref{fig:moment}, left panel) and  a flux measurement from 3.6 $\mu$m observations with the Infrared Array Camera (IRAC) onboard the {\it Spitzer} satellite  \citep[Table \ref{tab:measures}, ][]{Bouquin2018} we can calculate a distance to the galaxy with:\\
\begin{multline}
{\rm M_{bar}} ={\rm D^2} \times\left({\rm M/L}\times10^{-0.4(m_{3.6}-{\rm M}_{\odot 3.6})-2}+\right.\\ \left.2.36\times10^{-7} \times F_{\hi}\times1.33 \right)
\end{multline}
with D the distance in pc, M$_{\rm bar}$ the baryonic mass in M$_{\odot}$, M/L the mass to light ratio for the 3.6 $\mu$m IRAC band 1, $m_{3.6}$ and ${\rm M}_{\odot 3.6}$ the galaxy's apparent magnitude and the Sun's absolute magnitude in the same band, respectively, and $F_{\hi}$ the \hi\ flux in Jy \kms. Following \cite{Lelli2019} we use a factor 1.33 in order to account for the presence of helium and use M/L = 0.5 for the mass to light ratio in the infrared. Finally, as our $m_{3.6}$ is in the AB-magnitude system, we use ${\rm M}_{\odot 3.6}$ = 6.08 \citep{Willmer2018}. In this calculation we ignore the contribution of molecular hydrogen because \cite{Zabel2019} merely found upper limits for the galaxy and because the BTFR we are using was derived under the same assumption. \\
\begin{table}
    \centering
    \begin{tabular}{@{} lll @{}} 
       \multicolumn{3}{c}{ESO\,358-60}\\
       \hline
       \hline
	Model&M$_{\rm bar}$&Distance\\
 & ($\times 10^{8}$ M$_{\odot}$) & (Mpc)\\
       \hline
       Best fit& 4 $\pm$ 3 &9.5 $\pm$ 2.6 \\
       Single disk& 4  $\pm$ 3 &9.4 $\pm$ 2.5 \\
       Rising RC& 8 $\pm$ 4  & 13.7 $\pm$ 3.7 \\
       {\sc pyFAT}& 3 $\pm$ 2& 8.8 $\pm$ 2.4\\ 
       Hubble flow (CMB)$^{a}$& -  &10.6 $\pm$ 0.8 \\
       Cosmicflows-3 calculator$^{b}$& -  &10.1 \\
 \hline
    \end{tabular}
     \caption{Distances as derived from the fitted RC by estimating its baryonic mass through the BTFR. a) As listed on the NASA Extragalactic Database. b) \cite{Kourkchi2020} }
     \label{tab:distances}
 \end{table}
\indent The distances obtained for the various models are listed in Table \ref{tab:distances}. This table also shows the distance expected from the systemic velocity were the galaxy on the Hubble Flow and from the Cosmicflows-3 calculator \citep{Kourkchi2020}. It is obvious from these numbers that the BTFR distances are consistent with these independent distance estimates. In order to determine whether ESO 358-60 could be a likely member of the Fornax cluster at these distances, we estimate a lower boundary of the cluster by taking the minimal estimated distance \citep[D$_{\rm min}$ = 20 $-$ 1.7 Mpc = 18.3 Mpc, ][]{Blakeslee2009} and subtract twice the virial radius  \citep[R$_{\rm vir}$ = 0.7 Mpc, ][]{Drinkwater2001} from it. This leads  to a lower boundary for the Fornax cluster of 16.9 Mpc and we see that all the BTFR distances  fall significantly short of this distance. Only our upper distance limit, where the RC is forced to keep on rising, is consistent with the galaxy being part of Fornax at the upper ranges of the error bar.\\
\indent To ensure the applicability of the published BTFR to our galaxy, we compare several of its properties to a sub sample of the SPARC galaxies that have similar \vflat. These galaxies are highlighted in blue in Fig. \ref{fig:TF} and are defined by the box that can be drawn around ESO\,358-60 by setting the top right corner as the location where the rising RC model falls on the BTFR, i.e. the red box in Fig.\ref{fig:TF}. We will refer to these galaxies as the SPARC-sub sample in what follows.\\ 
\indent Fig. \ref{fig:RCs} compares the RCs of the SPARC-sub sample (gray lines) to  ESO\,358-60 at the best fit distance (red and blue lines indicating the receding and approaching sides, respectively) and the same at the distance of Fornax (pink and light blue lines). It is obvious from this comparison that to be at the distance of Fornax the RC of ESO\,358-60 would have to be extremely slowly rising, i.e. ESO\,358-60 would have a very shallow matter distribution in the inner parts. Even the distance of 9.4 Mpc appears on the high side in this comparison as the best fit RC rises slower than any of the curves in the SPARC-sub sample. The steepness of a RC, however, is also determined by the central  distribution of mass  \citep{Garrido2005, Lelli2022}. The central surface  brightness of ESO\,358-60 \citep[$\mu_{3.6}$ = 23.5 mag arcsec$^{-2}$, ][]{Bouquin2018} falls on the lower end of the distribution compared to the other galaxies in Fig. \ref{fig:RCs} \citep[SPARC-sub $\mu_{3.6}$ = 23.7 - 21.0 mag arcsec$^{-2}$, ][]{Lelli2016}. Hence we would expect ESO\,358-60  to have a RC that is relatively slowly rising compared to the SPARC-sub sample.  \\
\indent The stellar mass of the galaxy at 9.4 $\pm$ 2.5 Mpc, as derived from the 3.6 $\mu$m flux, is 3 $\pm\ 2 \times 10^7$ \msun\ and the \hi\ mass is 3 $\pm\ 1 \times 10^8$ \msun. The significant errors on these masses come predominantly from the uncertainty on the distance.  
The stellar and \hi\ masses of the SPARC-sub sample range from 0.6 - 17.6 $\times\ 10^7$ \msun\ and  0.5 - 4.1 $\times\ 10^8$ \msun, respectively.
This places the stellar mass of ESO 358-60 on the low end of SPARC-sub sample at the best fit BTFR distance but at the high end at the distance of Fornax (M$_{\rm star\ Fornax}$ = 12.3 $\times\ 10^7$ \msun).  
The \hi\ mass at the best fit distance on the other hand corresponds to the mean \hi\ mass in the SPARC-sub sample whereas it would be twice the maximum \hi\ mass in this sample if ESO\,358-60 were at the distance of Fornax (M$_{\rm \hi\ Fornax}$ = 11.5 $\times\ 10^8$). As the stellar and \hi\ mass scale in the same way with distance the galaxy certainly has a high \mhi$/$M$_{\rm star}$ ratio of 9.6. For the SPARC-sub sample this ratio ranges from 0.5-13.0. For ESO 358-60 \cite{Su2021} find a stellar mass, re-scaled to 9.4 Mpc, of 1.3 $\times\ 10^7$ from the $g'$, $r'$, $i'$ colors in the FDS. As this is even lower than the mass we derived from the 3.6 $\mu$m flux, we are convinced that the galaxy truly has a small stellar component.\\
\indent In order to further investigate whether ESO\,358-60 appears more as a field dwarf than a Fornax dwarf we compare the $g'-r'$ color of ESO\,358-60 to the colors of dwarfs in the Fornax cluster \citep{Venhola2018} and in the field as found in the Canes Venatici region \citep{Kovac2007, Kovac2009}. Fig. \ref{fig:CM-diagram} shows a color-magnitude diagram of the two samples and ESO\,358-60 at the best fit distance and at the distance of Fornax. Compared to the Fornax dwarfs  ESO\,358-60 is an outlier due to its blue colors, but compared to the dwarfs in the Canes Venatici region its color falls within the scatter of these galaxies. This is true regardless of which distance is considered.\\
\indent In our literature search on the galaxy we found one more redshift independent distance to ESO\,358-60. The galaxy was included into the Cosmicflows-2 survey \citep{Tully2013} who place the galaxy at 15.7 $\pm$ 3.1 Mpc based on an $I$-band TF-relation. The analysis presented in this paper provides several  improvements on the Cosmicsflows survey when considering individual galaxies. The most important of these improvements is that we include the \hi\ mass into the relation, i.e. we use the BTFR and not simply the TF-relation. The BTFR provides a tighter relation, especially for low mass galaxies \citep{McGaugh2000, Schombert2020} where the gas mass is as or more significant than the stellar mass. For such galaxies a relation based merely on optical light/stellar mass will overestimate the stellar contribution and thus yield a higher distance. This problem is exacerbated by the fact that \cite{Tully2013} have predominantly calibrated their relation on more massive galaxies (See their Fig. 8). In ESO 358-60 the \hi\ mass is $\sim$10 times the stellar mass and indeed if we repeat our analysis for \vflat\ and stellar masses instead of baryonic masses we find a distance of 13.7 $\pm$ 4.9 Mpc. Our distance is further confirmed by the Cosmicflows-3 calculator \citep{Kourkchi2020} whose calculated distance corresponds to our distance at the location and systemic velocity of ESO\,358-60.  As such we think no further explanation is required to account for the $\sim$ 6 Mpc difference with the \cite{Tully2013} result. \\  
\indent At the best fit distance of  9.4 Mpc the galaxy would be closer to us compared to what we would expect from its systemic velocity assuming a Hubble flow. The difference implies a peculiar velocity toward the Fornax cluster $\sim$ 100 \kms, indicating that the galaxy might feel the pull from the cluster and might eventually fall into it. If our best fit distance is still overestimating the distance to ESO\,358-60 the implied speed towards the cluster would increase simultaneously with its distance from the cluster. If this peculiar velocity is due to the gravitational attraction of the Fornax overdensity this would make it unlikely to be at a significantly lower distance. \\
\indent From the BTFR we deduce a distance to ESO\,358-60 of 9.4 $\pm$ 2.5 Mpc, placing it well in the foreground of the Fornax cluster. Such a distance fits  not only better with \vflat\ but also with the overall RCs and \hi\ content of low mass galaxies and the fact that the galaxy appears undisturbed and reasonably symmetric. \\
\section{Radial Motions}\label{RadialMotions}
Radial motions, also known as streaming motions, are likely to be a crucial element in the overall evolution of spiral galaxies. Inflows to the centers of galaxies can feed the central massive black hole \citep{Vollmer2008, Combes2023}, explain metallicity gradients \citep{Lacey1985, Sharda2021} and can in some cases even lead to bar-driven spiral arms \citep{Sanders1976, Sellwood2022}.  Foremost, inflows onto the inner disk are considered as an important candidate to sustain star formation in spiral galaxies \citep{Ho2019, EWang2022}. As there is not enough neutral gas available in the intergalactic medium (IGM) to feed the \hi\ in the inner disk directly from cosmological accretion \citep{Kamphuis2022}, such inflows might be required to transport gas from the outer disk to the star forming disk.\\
\indent Mechanisms resulting in large scale radial outflows have theoretically received less interest despite the fact that they have been unambiguously detected in several galaxies \citep{Schmidt2016}. Outflows due to supernova explosions are expected to occur with a preferred vertical direction \citep{Shapiro1976, Bregman1980, Norman1989} due to the steeper vertical density gradient. Indeed when holes, associated with supernova explosions, have been observed in \hi\ the ordered combined flow is in the vertical direction \citep{Boomsma2008} but the radial motions are circularly arranged around the center of the hole \citep{Kamphuis1991} creating random motions on the scale of the galaxy. Small scale radial outflows, as well as inflows, will occur along the spiral arms in disk galaxies \citep{Kalnajs1973} but we failed to find a theoretical background for larger scale outflows in the literature.\\
\indent One can imagine that the tidal torques of an interaction or the deposition of a large quantity of high angular momentum gas in the inner disk can drive a large scale outflow. Furthermore, additional pressure from a developing magnetic field driven by a magnetized turbulent dynamo \citep{Beck2012} can lead to an outward pressure on the gas. However, this depends fully on the details of the emerging magnetic fields and gas distribution in the galaxy \citep{Elstner2014}. Such mechanisms need further investigation, which is beyond the scope of this paper, in order to detail their net effects on the gas kinematics. Overall, however, there is enough evidence pointing to radial gas flows as an integral part of the  evolution of spiral galaxies.\\
\indent Typically, observed radial motions in \hi\ are of  the order of a few \kms \citep{Wong2004,Schmidt2016, diTeodoro2021} but can reach velocities of a few tens of \kms \citep{Fraternali2002}. When fitted with 2D tilted ring methods the determined radial motions are highly dependent on the underlying orientation of the model \citep{Schmidt2016} and they are partially degenerate with the PA \citep{Spekkens2007,EWang2023,SylosLabini2023}. As such they are typically difficult to identify, especially in the outer disk.\\
\indent ESO\,358-60 shows very clear radial motions in the center of the galaxy. They are undeniably present in the velocity field (Fig. \ref{fig:moment}) and also the 3D model is significantly improved by including radial motions (Fig. \ref{fig:parameters}). The motions appear to be mostly confined inside the high density gas accompanying the high density disk. The largest radius where we detect radial motions is slightly beyond D$_{\hi}$ (See Fig. \ref{fig:parameters}). The amplitude of the  motions in the model peaks at $\sim$ 15 \kms  at a radius slightly below 100\arcsec. The motions then continue all the way inward to the center at an approximately constant level around 10 \kms.\\
\indent These motions imply a peak mass flow rate, calculated by multiplying the \hi\ mass in a ring with its radial speed, of 0.5 \msun\ yr$^{-1}$ at R =  $\sim$ 100\arcsec. When we compare this to the star formation rate (SFR = 0.007 \msun\ yr$^{-1}$, see Table \ref{tab:measures}), such a flow would be adding more than enough gas  to the inner disk to sustain the SFR if the flow was inward. Unfortunately, without further observations or modeling we cannot establish the direction of the radial motions. To establish the direction of the radial motions we need to know the near and far side of the galaxy, or in other words we need to know the direction of rotation. Often, this can be established from auxiliary data through the presence of the dust lane or the winding of the spiral arms but  
 ESO\,358-60 lacks visible spiral arms or a prominent dust lane. As such, currently we see  no obvious path for determining, or even assuming, an orientation or rotational direction of the galaxy. Hence, we leave the interpretation of the radial motions in ESO\,358-60 for a future publication. \\ 
\section{Summary \& Conclusions}\label{Conclusions}
Due to a lack of anomalous gas in its surroundings and its very regular neutral hydrogen disk, the galaxy ESO\,358-60 stands out from other \hi\ detections in the MFS \citep[Fig. \ref{fig:overview},][]{Serra2023}. This leads us to create a detailed model of the \hi\  distribution in the galaxy to parametrize its disk in order to understand its regularity. By fitting a TRM in 3D to the \hi\ data of the MFS we are able to extract the kinematical and spatial parameters and obtain a detailed picture of the \hi\ distribution in the galaxy. The \hi\  distribution displays a significant, but still almost symmetric,  line of sight warp that starts around D$_{25}$ and shows a maximum shift of $\Delta i \sim 20^{\circ}$. Furthermore, the galaxy displays radial motions all over the observed optical disk and these peak between D$_{25}$  and D$_{\hi}$. The TRM further confirms the regularity of the disk. \\  
\indent  The most straightforward explanation for the regularity of the \hi\ disk, compared to other \hi\ detections in the MFS, is a lack of interaction which Fornax' ICM. From our TRM we establish a rotational velocity for ESO\,358-60 of \vflat\ =  48.1 $\pm$ 1.4 \kms. This translates to a redshift independent distance of 9.4 $\pm$ 2.5 Mpc for ESO\,358-60 when using this rotational velocity to place the galaxy on the BTFR derived from \vflat\  \citep{McGaugh2000, Lelli2019}. Comparing other properties of the galaxy, such as \hi\ mass or central stellar surface brightness, to galaxies from the SPARC database \citep{Lelli2016, Lelli2019} with similar \vflat\ leads to the conclusion that at our derived distance ESO\,358-60 is a normal dwarf Irregular with a  stellar mass and central surface brightness at the low end of the distribution. Had the galaxy been at 20 Mpc, the distance of Fornax, it would have to have an extraordinary amount of \hi\ compared to other galaxies with similar \vflat\ in the SPARC sample. The optical color of the galaxy matches a sample field galaxies better than the other dwarfs found in Fornax. However, this is true regardless of the distance to ESO\,358-60.\\
\indent We conclude that ESO\,358-60 at the retrieved redshift independent distance is more than $\sim$10 Mpc in front of the Fornax cluster. As such it is a field Irregular galaxy that, at best, is starting to feel the gravitational attraction of the Fornax cluster. Our modeling shows that the galaxy contains a significant line of sight warp as well as radial motions in its inner disk whose origins remain elusive for now. \\
\begin{acknowledgements}
We thank Prof. J. English for developing a perceptually correct velocity field color map for a white background. We thank F. Loi for useful advice and discussions on the ICM density of the Fornax cluster. We would also like to thank two anonymous referees who helped to improve the paper through their careful reading and useful comments. The work at RUB is partially supported by the BMBF project 05A23PC1 for D-MeerKAT. The MeerKAT telescope is operated by the South
African Radio Astronomy Observatory, which is a facility of the National Research
Foundation, an agency of the Department of Science and Innovation. We
acknowledge the use of the Ilifu cloud computing facility - www.ilifu.ac.za, a
partnership between the University of Cape Town, the University of the Western
Cape, the University of Stellenbosch, Sol Plaatje University, the Cape Peninsula
University of Technology and the South African Radio Astronomy Observatory.
The Ilifu facility is supported by contributions from the Inter-University Institute
for Data Intensive Astronomy (IDIA - a partnership between the University of
Cape Town, the University of Pretoria and the University of the Western Cape), the Computational Biology division at UCT and the Data Intensive Research
Initiative of South Africa (DIRISA). This project has received funding from the European Research Council (ERC) under the European Union’s Horizon 2020 research and innovation programme (grant agreement no. 679627, “FORNAX”; and grant agreement no. 882793, “MeerGas”). The data of the MeerKAT Fornax Survey are reduced using the CARACal pipeline, partially supported by
ERC Starting grant number 679627, MAECI (Italian Ministry of Foreign Affairs and International Cooperation) Grant Numbers PGR ZA23GR03 (RADIOMAP) and PGR ZA18GR02 (RADIOSKY2020), DSTNRF
Grant Number 113121 as part of the ISARP Joint Research Scheme, and
BMBF project 05A17PC2 for D-MeerKAT. Information about CARACal can
be obtained online under the URL: https://caracal.readthedocs.io. This research has made use of the NASA/IPAC Extragalactic Database (NED), which is funded by the National Aeronautics and Space Administration and operated by the California Institute of Technology. This research has made use of the VizieR catalogue access tool, CDS,
 Strasbourg, France (DOI : 10.26093/cds/vizier). The original description 
 of the VizieR service was published in 2000, A\&AS 143, 23\\
\end{acknowledgements}
\bibliographystyle{aa}
\bibliography{references}
\end{document}